\documentclass[12pt]{iopart}
\usepackage{graphicx}
\begin{document}
\title{Electrically induced tunable cohesion in granular systems}
\author{Jean-Fran\c cois M\'etayer$^{1,2}$, Patrick Richard$^1$, 
 Alain Faisant$^1$ and Renaud Delannay$^1$}
\address{$^1$
Institut de Physique de Rennes,  Universit\'e de Rennes 1, UMR CNRS 6251,
B\^at 11A, Campus Beaulieu, F-35042 Rennes Cedex, France
}
\address{$^2$ present address : Max-Planck-Institut f\"ur Dynamik und Selbstorganisation
Bunsenstr. 10,
D-37073 G\"ottingen,
Germany}
\ead{jean-francois.metayer@ds.mpg.de, patrick.richard@univ-rennes1.fr, alain.faisant@univ-rennes1.fr and renaud.delannay@univ-rennes1.fr}

\begin{abstract}
Experimental observations of confined granular materials in the presence of an electric field that induces cohesive forces are reported. The angle of repose is found to increase with the cohesive force. A theoretical model for the stability of a granular heap, including both the effect of the sidewalls and cohesion is proposed. A good agreement between this model and the experimental results is found. The steady-state flow angle is practically unaffected by the electric field except for high field strengths and low flow rates. 
\end{abstract}

\vspace{2pc}
\maketitle
\def \Vec#1{\overrightarrow{#1}}

\section{Introduction}

One striking property 
of a granular system is the ability to form a pile with a non-zero angle with the horizontal.
This angle, called the angle of repose $\phi_r$, is influenced by many factors such as the grain 
 roughness, polydispersity~\cite{Goujon2003} and confinement~\cite{Courrechdupont2003}.
 Cohesive forces~\cite{Restagno2004,Nowak2005,Taylor2008,Mason1999} also have an important influence on the angle of repose.
A recent paper~\cite{Mersch2010} has pointed out that an electric field may lead to cohesive forces within a granular pile.
This is an interesting issue for many practical applications since the cohesive forces can be almost instantaneously introduced into (removed from) the system by switching on (off) the electric field. 
Understanding the effect of an electric field on a granular pile is also of fundamental interest. For example, it has been reported that
ancient sediments had a larger angle of repose than present ones. 
A possible explanation~\cite{Kanagy1994} is the 
modification of Earth's electromagnetic field as well as the modification of electrical charging of grains due to climatic changes. 
The electric field has also an influence on blowing sand and dusty phenomena on Earth, Mars, the Moon and asteroids~\cite{Renno2008} and on the formation of geological patterns~\cite{Shinbrot2006}.
Moreover, in the majority of the works related to granular flows (see~\cite{GDRMIDI2004,Delannay2007} and references therein) the influence of electric forces is neglected. Although this assumption is reasonable for large~\cite{Bi2005, Bi2006} or metallic grains~\cite{Azanza1999}, it probably cannot be neglected for small plastic or glass beads as recently demonstrated in \cite{Pahtz2010}. Imposing an external well-controlled electric field is a good way to measure the influence of the electric forces on granular flows.\\
Mersh et al.~\cite{Mersch2010} list the electrically induced interactions a packing of glass beads can experience and compare their relative magnitudes. These interactions are: (i) 
dipole/dipole interactions due to the polarization of the grains,
(ii) electrostatic force,
(iii) interaction between two charged grains, and 
(iv) electroclamping force.
We refer the reader to the work~\cite{Mersch2010} (and references therein) for details. For the majority of granular systems, only the electroclamping force~\cite{Dietz1978b} is significant. We will show below that this is indeed the case for our system. Let us recall that 
when a packing of glass beads is confined between two electrodes a small electric current can pass through the system due to the presence of absorbed moisture. The density of  this current increases at the contacts between the grains leading to a local electric field at the surface that can be several orders of magnitude higher that the applied electric field. Dietz and Melcher~\cite{Dietz1978b} propose the following semi-empirical formula: $F^{i,j}_d = 0.415\pi\varepsilon_0 d^2 E_{sat}^{0.8}(E\cos\alpha)^{1.2},$
where $d$ is the grain diameter,  $E$ the electric field, $\alpha$ the angle between the electric field and the normal to the contact  and $E_{sat}=3\times10^6\mbox{ Volts/m}$ is the nominal electric field for dielectric
breakdown in air (the effect of temperature, humidity, pressure\ldots on this quantity is neglected).\\ 
An electric field may therefore lead to the presence of cohesive forces within  granular systems that are physically different from the adhesive forces due to the presence of liquid bridges~\cite{Mason1999,Nowak2005}. Furthermore, the strength of these forces is set by the strength of the electric field, which means that it is possible to tune cohesion in a granular system by modifying the strength of the electric field. Understanding the effect of an electric field on the stability of a granular pile will then shed light on  the stability of cohesive granular heap. 
Note that cohesion can also be induced by a magnetic field~\cite{Peters2004,Lumay2007,Taylor2008} but in this case a residual magnetization might be a problem. {Moreover this is possible only for magnetic grains whereas an electric field can induce cohesion within any granular system as long as a small electric current can pass through the system due to absorbed moisture (see below).} 


Our goal is to quantify, using precise and systematic  measurements of the angle of repose in the presence of an electric field,  the influence of this field both on the static and on the dynamic properties of granular systems.
One of the main results is that an electric field allows for the precise control of the angle of repose of the heap
within a range that depends on the width of the pile. It is also possible to control the start or the stop of a flow on the pile by switching  the electric field off or on. A theoretical model capturing the effect of the electrically induced cohesion as well as the effect of the confinement is presented. The effect of an electric field on the dynamic properties of the flow is found to be weak.

\section{Experimental set-up}
\begin{figure}[htbp]
\begin{center}
\includegraphics*[width=10cm]{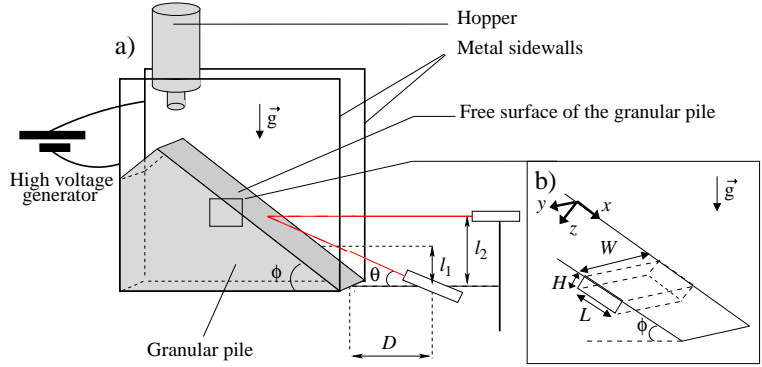}
\caption{a) Sketch of the experimental set-up. Parallel plates are made of Duralumin. Two lasers are used to measure the angle of repose of the heap (see text). b) Sketch of the pile for the calculation of the equilibrium of a thin layer.}\label{manip}
\end{center}
\end{figure}

Experiments are performed using a quasi-two-dimensional cell. It consists of two vertical, flat and parallel metallic plates ($70~\mbox{cm} \times 70~\mbox{cm}$), as sketched in figure~\ref{manip}(a). The cell is totally closed at its bottom and on its left, so that the beads can form a pile. This configuration enables us to control the width $W$ between the plates. The granular material we used is made of glass beads of diameter $d=(500\pm100)~\mu m$ and of density $\rho_g = 2500\mbox{ kg.m}^{-3}$. Note that these grains have been used in a previous study on confined flow rheology~\cite{Richard2008}.
In our experiments $W$ ranges from $8d$ to $73d$.
The humidity ($45\%$) and the temperature ($23^\circ\mbox{ C}$) are controlled.

To obtain a static electric field in the cell we used a high voltage generator (from $V = 0\mbox{ Volt}$ to $3000\mbox{ Volts}$). The corresponding electric field is, by definition, $E=V/W$. One of the metallic plates is maintained at a potential of $0\mbox{ Volts}$ during all the experiments, and  the potential of the other plate is freely adjustable.
To form a pile between the two sidewalls we use the following protocol. First we impose a voltage $V$ (and thus an electric field) between the sidewalls. 
Then, grains are continuously poured between the two sidewalls from a hopper. 
A static heap slowly
forms by trapping grains at its top~\cite{Taberlet2004b}. After a transient,
the growth of the heap stops and the flow at
its surface reaches a steady state. Then, the feeding of grains is cut off, and the angle of the remaining heap is measured once all the grains in the cell are static.  
In order to measure the angle of the heap we used two lasers as sketched in figure~\ref{manip}a). The first one is placed on a rotating axis whose height is the same as 
the lowest point of the cell. The distance between the laser and the exit of the cell, $D$, is
fixed. The beam is always parallel to sidewalls and is inclined from the horizontal 
with  an angle $\theta$.  
The second laser is placed horizontally and its height is adjustable. The two beams are reflected on the pile. We adjust the height of the second laser in order to superimpose the two light reflections on the pile. By measuring $l_1$ and $l_2$, respectively the height between the first laser and the reflection of its beam on the cell and  the height of the second laser (see figure~\ref{manip}), the angle of the heap can be calculated  using equation~(\ref{angle}) with a precision of $\pm 0.1^\circ $:
\begin{equation}
\label{angle}
\tan\phi = \frac{l_2.l_1}{D(l_2-l_1)}.
\end{equation}
\section{Influence of the electric field on the stability of a pile}
\subsection{Electrically induced interactions}
The presence of a voltage difference on the sidewalls leads to extra forces on grains. 
Those interactions~\cite{Mersch2010} are listed below  and their maximal magnitude (obtained with the largest value of the electric field used in our experiments $E=2\times 10^5\mbox{ Volts/m}$) is compared to the weight of a grain. We recall that in our experiments the grains are glass beads (density $\rho_g=2500\mbox{ kg.m}^{-3}$). Their diameter is $d$ and their relative permittivity $\varepsilon_r$. 
\begin{itemize}
\item Dipole/dipole interactions $F_a$ due to the polarization of the grains. 
The ratio of $F_a$ to $mg$ is approximately $2\times 10^{-2}$ (see~\cite{Mersch2010} for calculation). 
\item Electrostatic force : a grain $i$ of charge $q_i$ immersed in the applied field can experience a force $\Vec{F_b^{i}}=q_i\Vec{E}$. 
In order to measure the mean charge per grain $q_{mean}$, $N_g$ grains are poured in a Faraday cage. In contact with  one of the electrodes of a capacitor (whose capacity $C$ is known) the grains charge, by influence, the capacitor. The measurement of the electric potential $U_{cap}$ between the electrodes is linked to $q_{mean}$ through : $q_{mean}=U_{cap}/(N_gC)$.   
We obtained $q_{mean}=1\times 10^{-14}\mbox{ C}$, therefore the maximal electrostatic force $F_b$ (for $E=2\times 10^5\mbox{ Volts/m}$)) is $2\times 10^{-9} N$. The ratio of $F_b$ to $mg$ is then $4\times 10^{-7}$.
\item Two charged grains can also interact~\cite{Feng2000}. 
 If the charge $q_i$ and $q_j$ of these grains are both equal to the saturation charge, the ratio of the corresponding force $F_c$ to $mg$ is approximately $4\times 10^{-3}$.
\item The electroclamping force.
 Following the semi-empirical formula proposed by Dietz and Melcher~\cite{Dietz1978b}: $F_d^{i,j} = 0.415\pi\varepsilon_0 d^2 E_{sat}^{0.8}(E\cos\alpha)^{1.2},$
where $\alpha$ is the angle between the electric field and the normal to the contact. If $\alpha=0$ the ratio of $F_d$ to $mg$ is $0.6$.
\end{itemize}
The only force which is not small compared to the weight of a grain is the electroclamping force
$F_d=(F_d^x,F_d^y,F_d^z)$. This force is attractive and will increase the cohesion of the packing. 
The grain-sidewall interactions can be determined by using the method of images. Each grain in contact with the sidewall has an image located symmetrically about the sidewall and with an opposite charge. Therefore the electroclamping force for a grain-sidewall interaction is
$F_d^{i,w} = 0.415\pi\varepsilon_0 d^2 E_{sat}^{0.8}E^{1.2}=k_DE^{1.2}.$ 
\subsection{Experimental results}
\begin{figure}[htbp]
\begin{center}
\includegraphics*[width=8cm]{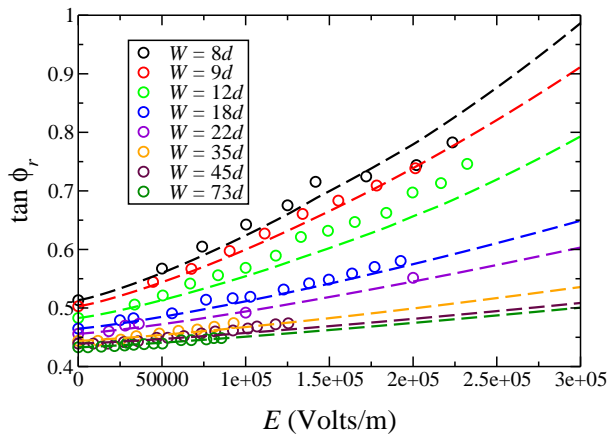}
\caption{(color online) The measured angle of repose (symbols) increases with the electric field. This effect, which is stronger as the gap between side walls decreases, is very well described by equation~(\ref{anglereposelec}) (dashed lines).}
\label{data}
\end{center}
\end{figure}

For a width $W$ between the plates varying from $8d$ to $73d$, we  measure the angle of repose 
without an electric field ($E=0$), $\phi_{r}(0)$ and with an electric field $\phi_r(E)$ using the protocol described above. 
The results are presented in figure~\ref{data}.
The data reported in this figure clearly demonstrate that both the confinement and the electric field have an effect on the angle of repose. First, without an electric field the angle of repose increases as $W$ decreases. This result is in agreement with well known results~\cite{Liu1991,Courrechdupont2003,Boltenhagen1999}. Second, in the presence of an electric field, the angle of repose can be increased by more than $10^\circ$ (for $W$ equal to $8d$ and $E$ equal to $2.2\times 10^5\mbox{ Volts/m}$).  It is therefore necessary to introduce cohesion and confinement effects in any model that aims to quantitatively predict the stability of confined and cohesive granular systems.\\
%
It should be pointed out that the angle of repose, for a given width $W$,  depends on the current value of electric field $E$ and not on its history. In other words, for a given value of $E$, the angle of repose will be the same if $E$ is reached directly from $0\mbox{ Volt}$, or if  several measurements for smaller value of $E$ are made before. Starting from the repose angle at a given $E$,  the angle will reach the value $\phi_{r}(0)$, independently of the previous value of $E$ when the electric potential is switched off. By switching on the electric potential before the heap reaches $\phi_{r}(0)$, the cell emptying will stop.
\subsection{Model}
\begin{figure}[htbp]
\begin{center}
\includegraphics*[width=8cm]{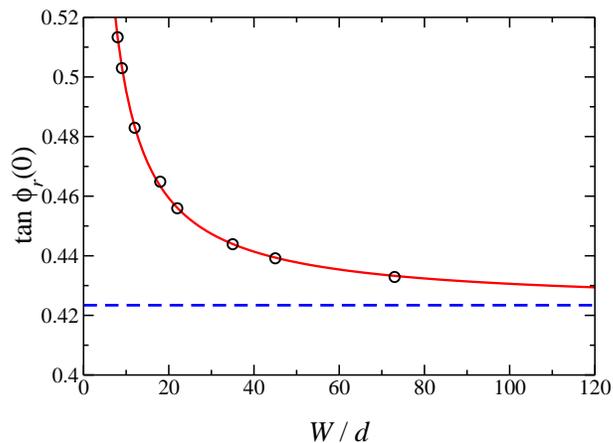}
\caption{
(color online) In the absence of an electric field the angle of repose depends strongly on the width between the sidewalls: a smaller width leads to a larger angle of repose. The red line corresponds to the fit $\tan \phi_r = \tan\phi_r^\infty + {\mu_W H}/{W}$ and the blue dashed line to $\tan\phi_r = \mu_B=\tan\phi_r^\infty$.}\label{fig:theta_0}
\end{center}
\end{figure}

We now derive a model that is able to capture the evolution of the angle of repose in the presence of an applied electric field. Note that this model differs significantly from the one presented by Robinson and Jones~\cite{Robinson1984b} which takes into account the effect of the electric field (on the grain-grain interactions as well as the grain-wall interaction) but not the confinement. As mentioned above, the width $W$ between the sidewalls is known to have a strong influence on the angle of repose of the heap: it increases for decreasing $W$. This effect is even more important for cohesive grains~\cite{Nowak2005}.  Following~\cite{Courrechdupont2003} we can derive a simple relation between $\phi_r$ and $W$. 
Let us consider a slab of material of  width $W$ parallel to the free surface of the pile (figure~\ref{manip}(b)). A surface avalanche starts at a given depth $H$ when the equilibrium is broken in the direction tangent to the free surface of the pile. 
In the absence of electric field, the slab is only subjected to its weight as well as to the forces exerted at the two sidewalls and at the base of the slab. According to the Coulomb friction law, the
shear forces on the two sidewalls and at the base of the slab
are then respectively
\begin{equation}
F_W=2\mu_W N_W, 
\end{equation}
and 
\begin{equation}
F_B=\mu_B N_B = \mu_B \rho g W L H \cos\phi_r,
\end{equation}
where $N_W$ and $N_B$ are respectively 
the normal force on a  sidewall and the normal force at the base,
$\mu_W$ and $\mu_B$ are respectively the effective friction coefficients at the sidewall and at the base,
and $\rho=\nu \rho_g$ is the density of the granular medium ($\nu$ is the packing fraction of the medium).
In the following we will assume that the effective friction coefficients $\mu_B$ and $\mu_W$ are constant (i.e. they do not depend  on $W$ or on $y$).
Assuming that the pressure is isotropic, we have $N_W=L\int_0^{H}p(h)dh$,
where $p(h)$ is the pressure at height $h$. 
The force balance along the $x$-direction, $mg\sin\phi_r = F_B+F_W$, can be written as
\begin{equation}
\tan\phi_r = \mu_B + \mu_W \frac{2N_W}{W\rho g L H \cos\phi_r}.\label{eqn:tan}
\end{equation}
From the previous equation, it can be seen that the effective friction coefficient
$\mu_B$ is equal to $\tan\phi_r^\infty$, the tangent of angle of the heap without walls (i.e.
$W\rightarrow \infty$). The previous equation can then be rewritten as 
\begin{equation}
\tan \phi_r = \tan\phi_r^\infty + \frac{B_m}{W},
\label{dupont}
\end{equation}
where   $B_m=2\mu_W{\int_0^{H}p(h)dh}/{p(H)}$.
The form of this previous equation is similar to the one derived
in~\cite{Courrechdupont2003,Courrechdupont2003b}.
If we assume that the pressure is hydrostatic we have $B_m=\mu_WH$ and
\begin{equation}
\tan\phi_r = \tan\phi_r^\infty + \mu_W\frac{H}{W}.
\label{eqn:SSHlike}
\end{equation}
In figure~\ref{fig:theta_0} we report the  angle of repose versus $W$ without an electric field. The red line, corresponding to $\tan \phi_r = \tan\phi_r^\infty + \mu_W H/{W}$, is in excellent agreement with the experimental data. 
From this equation we can measure 
the quantity $B_m=\mu_W H$ which is the characteristic length of the lateral wall effects.
From our experimental results for $E=0$ 
and for the different  $W$ and using equation~(\ref{dupont}), we are able to determine the values of $B_m$ and $\phi_r^\infty$. We obtained $B_m=0.716d$  and $\phi_r^\infty=22.94^\circ$.
The metal-glass static friction coefficient is between $0.5$ and $0.7$~\cite{CRC2008}, therefore it is reasonable to choose $H\approx d$. Moreover we will assume in the following that this height does not depend on the intensity of the electric field. This is reasonable as long as the cohesion force is not much greater than the weight of the grains. As mentioned above this is the case in our experiments since the maximum electric force is approximately $0.6 \mbox{ }mg$. Note that  if we assume that equation~{(\ref{eqn:SSHlike})} still holds just before the flow atop the heap stops, $H$ can be seen as the flowing layer height when the flow is about to stop. In this case $H\approx d$ is also reasonable for moderate cohesion.
We now have to introduce the effect of the cohesive forces due to the electric field in the bulk and at the sidewalls. 
Let $\sigma_E^{i,j}$ with $i,j=\{x,y\}$ and $\sigma_E^W$ be respectively the stress tensors components corresponding to the cohesive force $F_d$  in the bulk and the  normal stress at a sidewall corresponding to the cohesives forces $F_d$ exerced by the grains in contact with that sidewall. We assume that the stress normal to the surface is the sum of the hydrostatic pressure and the normal stress due to the electric force. Therefore the normal stresses at the base of the slab and on a sidewall are respectively  $P(z=H)+\sigma_E^{zz}$ and $P(z)+\sigma_E^{yy}+\sigma_E^W$, where $\sigma_E^W$ is the stress tensor corresponding to the grain-wall contact. 
The stability condition is then
$$
\rho g H L W\sin\phi_r =  \mu_BWL\left(\rho g H  \cos \phi_r+ \sigma_{E}^{zz} \right) \displaystyle{+2\mu_WL\int_{0}^{H}\left(P(z)+\sigma_{E}^{yy}+\sigma_{E}^{W}\right)dz,}
$$
which yields
\begin{equation}
\tan\phi_r = \displaystyle{\mu_B\left(1+\frac{\sigma_{E}^{zz}}{\rho g H \cos\phi_r}\right)}
 \displaystyle{+\mu_W\left(\frac{H}{W}+2\frac{\sigma_E^{W}+\sigma_{E}^{yy}}{{\rho g W\cos\phi_r}}\right)}\label{eqn:tantheta}.
\end{equation}
This equation links the repose angle and the depth of the failure.
Note that for a given $\phi_r$, two values of $H$ can be obtained. The lowest one corresponds to a failure in the bulk; the highest one  corresponds to a failure at the walls. This point will be addressed in a future paper~\cite{Richard2010}. 
To derive a relation between the cohesion force $F^d_w$ and the stress tensor at the sidewalls $\sigma_E^W$ we have to determine the number of grains in contact with a sidewall. Recently Suzuki et al.~\cite{Suzuki2008} showed that the presence of a  sidewall increases the packing fraction. In the bulk the packing fraction is close to $\nu_{bulk}=\rho/\rho_g=0.6$ whereas it is close to $\nu_{wall}=\rho_{wall}/\rho_g=0.8$ near the wall.  Therefore the number of grains in contact with a wall of area $LH$ is $4 \rho_{wall} LH/ (\rho_g\pi d^2)$. The corresponding normal stress is then
$$ \sigma_E^W = \frac{4 \rho \kappa }{(\rho_g\pi d^2)}F^d_w,$$
with $\kappa=\nu_{wall}/\nu_{bulk}\approx 4/3$.\\
To obtain an expression for the stress components $\sigma_E^{ij}$ in terms of
electric force $\Vec{F_d}$ and the branch vectors $\Vec{l}$ joining the
particle mass centers we use the formalism developed in~\cite{Babic1997}: $\sigma_E^{ij}=V^{-1}\sum_{C\in V} u(C) F_d^i(C) l^j(C)$, where $V$ is the control volume, $C$ denotes the contacts
in $V$ and $u(C)$ is the fraction of length $l(C)$ contained in $V$.
This requires the knowledge of the texture of the granular medium.
For the sake of simplicity we consider a mean field approach where the contact network is organized as a cubic lattice.
Each grain in the bulk has $N_c=6$ contacts, $N_{z}=N_{y}=N_{x}=2$ in each direction. This simplification leads to 
$\sigma_{E}^{ij}=2\delta_{ij}\times {3\rho{F^d_i}}/{\rho_g \pi d^2},$
i.e. to  $\sigma_{E}^{zz}=0$ and to $\sigma_{E}^{yy}={6\rho F^d_y}/{\rho_g \pi d^2}.$
Equation~(\ref{eqn:tantheta}) becomes
\begin{equation}
\tan\phi_r = \mu_B+\mu_W\left(\frac{H}{W}+\frac{4\left(2\kappa F^d+3F^d_y \right) }{\pi \rho_g g d^2 W\cos\phi_r}\right),\label{anglereposelec}
\end{equation}
which yields
\begin{equation}
\frac{W}{d}\cos\phi_r\Delta\tan\phi_r=  \frac{4\mu_W }{\pi \rho_g g d^3}\left(2\kappa F^d_w+3F^d_y \right),\label{eqn:scaling}
\end{equation}
with $\Delta\tan\phi_r=\tan\phi_r(E)-\tan\phi_r(0)$.
Note that in our case, $F_d^y=F_d^w=k_DE^{1.2}$.

\begin{figure}[h!]
\begin{center}
\includegraphics*[width=8.cm]{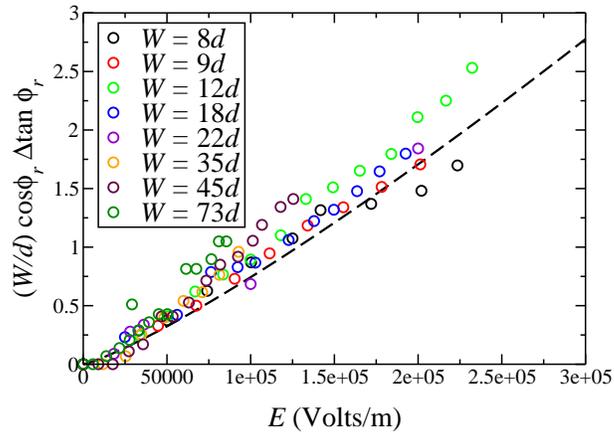}
\caption{(color online) Our model predicts that $(W/d)\Delta\tan\phi_r\cos\phi_r$ collapses onto a master curve when plotted against the electric field $E$. The symbols represent the experimental data and the dashed line represents the equation~(\ref{eqn:scaling}).}
\label{fig:scaling}
\end{center}
\end{figure}

We have shown in figure~\ref{data} the comparison between the experimental data and the model derived above. The agreement is very good, both the effects of 
$W$ and $E$ are correctly taken into account. 
The eletroclamping force and the confinement both facilitate the formation of arches in the granular system.
 So, keeping the electric field $E$ constant and increasing the gap between sidewalls should
 reduce the effect of the electrical field. This is confirmed by our experimental measurements
 (see figure~\ref{data}).
To test this model more carefully we also compare our data directly with equation~(\ref{eqn:scaling}) as reported in figure~\ref{fig:scaling}. 
Here again the agreement is found to be very good. A small deviation can be found for large  $W/d$ (the model slightly underestimate the experimental values). This is due to the simplified structure used  in our model that leads to $\sigma_{E}^{zz}=0$. For a real structure, $\sigma_E^{zz}$ is probably small but it is not exactly zero. 
\section{Influence of the electric field on Surface flows}
\begin{figure}[htbp]
\begin{center}
\includegraphics*[width=8.cm]{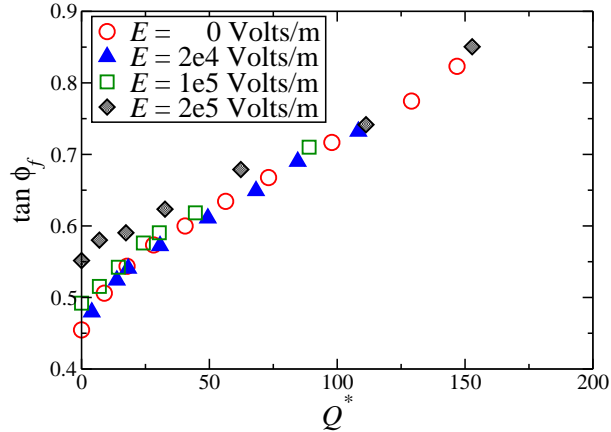}
\caption{(color online) Tangent of the angle of the flow as a function of the dimensionless flow rate.}
\label{dataec}
\end{center}
\end{figure}

Using the same apparatus, it is also possible to calculate the angle of a steady flow on the pile for different values of $E$. The hopper is now continuously filled with particles and its aperture precisely controls the input flow rate $Q$, defined as the mass of material entering the channel per unit of time and per unit of width $[Q = mass/(time . width)]$. When the system reaches a steady state, the input flow rate is equal to the output flow rate which is calculated using an electronic scale which weights the material falling out of the channel. The results obtained for a width $W$ equal to $22d$ are presented in figure~\ref{dataec}. In this figure $Q^*$ is the dimensionless flow rate defined as: $Q^* = Q/(\rho_g\nu_{bulk} d\sqrt{gd})$ with $\nu_{bulk}=0.6$. The increase of the tangent
of the angle of the free surface of the flow, $\tan(\phi_{f})$, with increasing $Q^*$ has already been the subject of many careful studies~\cite{Taberlet2003,Jop2005}. Here we focus on the variation of $\tan(\phi_{f})$ with the electric field. 
As reported before in this paper, the angle of a static pile (i.e. for a zero flow rate)
 increases with $E$.
The case of granular flows is more complex. As reported in figure~\ref{dataec}, the influence of the electric field is significant only for
low input flow rates.\\

For large  flow rates, the behavior of our system is no longer influenced by the electric field. The reason for this is straightforward.  As mentioned before, the electroclamping force is the only non-negligible electric force in our system. This interaction exists only if the grains (or if the grain and the wall) are in contact. 
A large flow rate leads to a large angle for the free surface~\cite{Taberlet2003}
and a relatively dilute flow~\cite{Taberlet2008}. Thus, close to the free surface, the neighborhood
of a given grain is dilute, the interactions between grains are mostly collisional and there is almost no enduring contacts.
The average number of contacts per grain is small and the electroclamping force
is thus negligible. 
\section{Conclusion}

We have reported experimental results on the effect of an electric field on the angle of repose, $\phi_r$, of a granular heap. We show that an electric field induces cohesive forces between two grains in contact, and between the sidewalls and the contacting grains. This electrically induced cohesion increases the stability of the pile. This phenomena has been studied for the width between the walls ranging from $8d$ to $73d$ and can lead to an increase in $\phi_r$ of more than $10^\circ$. A theoretical approach of the behavior of this angle is proposed, based on previous works~\cite{Courrechdupont2003,Mersch2010}, including the effect of the sidewalls and of the electric field.
 The angle of a steady flow can be increased by an electric field but this effect is only observed for low  flow rates.
\section{Acknowledgments}
We are indebted to Dr. J\'er\^ome Lambert for many fruitful discussions and advice on electrostatics in granular systems. 
We are grateful to Dr. Luc Oger and to Pr. J\'er\^ome Crassous for many valuable discussions. We thank J.R. Th\'ebault for technical assistance, Dr. Sean McNamara and Dr. Zeina S. Khan for a critical reading of the manuscript and Dr. Geoffroy Lumay for providing us with reference~\cite{Mersch2010} before publication.\\ This work is supported by the region Bretagne (CREATE grant : Sampleo) and by ANR grant (ANR-05-
BLAN-0273).\\


\bibliography{biblielec}
\bibliographystyle{unsrt}
 

\end{document}